\documentclass[fleqn,usenatbib]{mnras}
\usepackage[T1]{fontenc}
\usepackage[utf8]{inputenc}
\usepackage{pslatex}
\usepackage{graphicx}
\usepackage{floatrow}
\usepackage{subcaption}
\usepackage{caption}
\usepackage{amsmath}
\usepackage{hyperref}
\hypersetup{
    colorlinks=true,
    linkcolor=blue,
    filecolor=green,      
    urlcolor=blue,
}
\usepackage{siunitx}
\usepackage{gensymb}

\graphicspath{ {./images/} }
\title[GeV Emission Region in Jets of Blazars]{Locating the GeV Emission Region in the Jets of Blazars from Months-Timescale Multi-Wavelength Outbursts}

\author[Barat et al.]{
Saugata Barat,$^{1,2}$ \thanks{E-mail: saugatabarat500@gmail.com}
Ritaban Chatterjee$^{1}$
and Kaustav Mitra$^{3}$\\
$^{1}$Department of Physics, Presidency University, 86/1 College Street, Kolkata 700073, India.\\
$^{2}$Anton Pannekoek Institute for Astronomy, University of Amsterdam, Science Park 904, NL-1098 XH Amsterdam, the Netherlands.\\
$^{3}$Department of Astronomy, 52 Hillhouse Avenue, Steinbach Hall, Yale University, New Haven, CT 06511, USA.\\
}

\date{Accepted  Received ; in original form }

\pubyear{2022}

\begin{document}

\label{firstpage}
\maketitle

\begin{abstract}
It is well-known that the $\gamma$-ray emission in blazars originate in the relativistic jet pointed at the observers. However, it is not clear whether the exact location of the GeV emission is less than a pc from the central engine, such that it may receive sufficient amount of photons from the broad line region (BLR) or farther out at 1-100 pc range. The former assumption has been successfully used to model the spectral energy distribution of many blazars. However, simultaneous detection of TeV $\gamma$-rays along with GeV outbursts in some cases indicate that the emission region must be outside the BLR.  In addition, GeV outbursts have sometimes been observed to be simultaneous with the passing of a disturbance through the so called ``VLBI core,'' which is located tens of pc away from the central engine. Hence, the exact location of $\gamma$-ray emission remains ambiguous. Here we present a method we have developed to constrain the location of the emission region. We identify simultaneous months-timescale GeV and optical outbursts in the light curves spanning over 8 years of a sample of eleven blazars. Using theoretical jet emission models we show that the energy ratio of simultaneous optical and GeV outbursts is strongly dependent on the location of the emission region. Comparing the energy dissipation of the observed multi-wavelength outbursts and that of the simulated flares in our theoretical model we find that most of the above outbursts originate beyond the BLR at approximately a few pc from the central engine.   
\end{abstract}



\section{Introduction}
Blazars are the most common non-transient source of $\gamma$-ray emission in the sky \citep[][]{abdolahi20,hovatta19}. Blazars have a powerful jet pointed toward the observers and hence their spectral energy distribution (SED) is dominated by the relativistically beamed emission from the jet \citep[][]{urry1995}. However, the exact location in the jet at which $\gamma$-rays are produced and the emission mechanism involved are not well-constrained. In the so-called leptonic model, the GeV emission is produced by the inverse-Compton (IC) scattering of seed photons by the relativistic electrons in the jet \citep[eg.][]{sikora_1994,Coppi_1999,B_a_ejowski_2000,Dermer_1997,Chiang_2002,Arbeiter_2005}. The same electrons generate the radio to optical and sometimes UV-X-ray emission via the synchrotron process \citep[eg.][]{bregman_1981,urry_1982,impey_1988,marscher_2005}. The broad line region (BLR) is considered as a prominent source of the said seed photons. The SED is fit well by models in which the emission region is within the BLR and it is the dominant source of seed photons \citep[][]{blanford1995,hartman2001,Dermer_1997,sikora_1994,ghisselini_96}.
However, detection of $\sim$0.1 TeV $\gamma$-rays from some flat spectrum radio quasars (FSRQs) indicates that the emission region is outside the BLR. Otherwise, photons above 20 GeV would be absorbed due to the $\gamma-\gamma$ interaction with the BLR photons \citep[][]{costamate18,tavecchio2009,donea2003}. 

On the other hand, comparison of GeV variability with that at 15 GHz radio emission of hundreds of blazars have been used to conclude that the dominant GeV emission zone is typically located at a distance of several parsecs from the central engine \citep{Kramarenko2022}. Furthermore, GeV flaring has often been observed to be contemporaneous with a disturbance passing through the so called VLBI ``core'', which implies that the core may be a possible site of the GeV emission \citep[][]{jorstad13,marscher2016}. In some blazars, rotation of the optical polarization vector and subsequent flaring in GeV and optical as well as the passage of a bright radio knot through the core have been construed to imply that the core is the site of bright $\gamma$-ray emission \citep[][]{marscher2008,D_arcangelo_2009,Agudo_2011}. The core, in turn, has been shown to be located a few to tens of pc down the jet from the central engine, well beyond the upper limits of the BLR radius. In that case, contribution of seed photon from some other source, such as, the dusty torus is needed \citep[e.g.,][]{Sikora_2009}.  

Recently, \citet{georganopoulos2020} have used the energy density of the photon population which is up-scattered by the jet in a sample of more than sixty blazars to conclude that in most of them the GeV emission occurs near the dusty torus located more than 1 pc from the black hole. They studied a diagnostic called the ``seed factor'' which has been shown to depend on observational quantities such as the peak IC and peak synchrotron luminosity. From a distribution of seed factors from observed FSRQs, they concluded that the GeV emission region is likely to be located near the molecular torus region. \citet[][]{nalewajko14} used constraints obtained from collimation parameter, synchrotron luminosity and EC photon energy corresponding to efficient radiative cooling to derive limits on the location of the GeV emission region. In all the cases studied they also find that the region must be located beyond the BLR. \citet{cao2013} constrained the GeV emission region of 21 FSRQs by studying their SEDs. If the emission region is within the BLR, IC scattering occurs in the Klein-Nishina regime and the spectrum in the GeV region is expected to have a broken power-law shape whereas if the IC happens in the Thomson regime, the GeV spectrum has the same spectral index as the optical/infrared regime. The authors reported that for 16 out of the 21 blazars in their sample, they found evidence for the location of the emission region to be beyond the BLR. From the broadband SED properties of more than 100 low synchrotron peaked (LSP) blazars in the radio-Planck sample, \citet{arsioli2018} concluded that EC process with IR dominated seed photons is the dominant mechanism of GeV emission, which indicates that the emission region is near the torus and much beyond the BLR. Similar conclusion has been reached by \citet{kang2014} using another sample.

In the BL Lac type sources, in which the broad line region and torus are weak or absent, the only plausible source of seed photons is the jet itself. In some of those sources the SED is fit well by the synchrotron self-Compton (SSC) model, i.e., the synchrotron photons produced in the jet act as the seed photons for the IC process \citep[eg.][]{maraschi1992,Chiang_2002,Arbeiter_2005}. However, in some other cases in which a typical SSC model does not work well and even in some FSRQs, despite the presence of the BLR and torus, another source of seed photons that has been considered by some authors is a slow-moving sheath around the faster spine of the jet containing the higher energy electrons \citep{ghisselini2005,macdonald15,macdonald17}. This is also relevant in the cases in which the emission region is near the VLBA core, which may be located farther down the jet much beyond the torus. Observational evidence for the existence of such spine-sheath structure of blazar jets have been found by, e.g., \citet{aleksi14}. 

While the location of the emission region far beyond the BLR is consistent with the observation of TeV $\gamma$-rays sometimes it is difficult to create heavily Compton dominated flares in blazars without the supply of optical-UV seed photons from the BLR for up-scattering \citep[e.g.,][]{Ghisselini_2010}. \citet{bottcher2016} have shown that if the $\gamma$-ray emission region is located near the outer boundary of the BLR then the $\gamma-\gamma$ absorption by the BLR radiation field may not be significant. Furthermore, in some blazars a significant correlation have been found between changes in the GeV flux and some of the broad emission line flux such as MgII \citep[][]{chavushyan20, tavares13, Isler_2013, isler2015}, which indicates that some $\gamma$-rays are produced within the BLR. Hence, uncertainty in the location of GeV emission still remains. It is, of course, possible that GeV emission is generated at multiple locations in the jet, some of which are within the BLR and the rest are near the torus or even beyond that.

In this paper, we study simultaneous outbursts in GeV and optical wave bands in order to investigate the location of those emission. The GeV and optical variability are well-correlated in most cases in a large sample of Fermi blazars \citep[][]{majumdar_19, Liodakis_2019}. Therefore, if the flares are within a few days of each other we may assume they are co-spatial and being generated as part of the same ``event.'' Assuming they are being produced in the same event provides some additional information and with that we can put constraints on the source of seed photons, and in turn, constrain the location of the emission region. In {\S}2 we describe the data reduction, the analyses of GeV and optical data in order to identify simultaneous flares and their properties as well as the theoretical model we use to interpret the results of the analyses are presented in {\S}3. In {\S}4 we show the results and discuss their implications. 
\begin{figure*}
    \begin{subfigure}{.5\textwidth}
  \centering
  \includegraphics[width=1.0\linewidth]{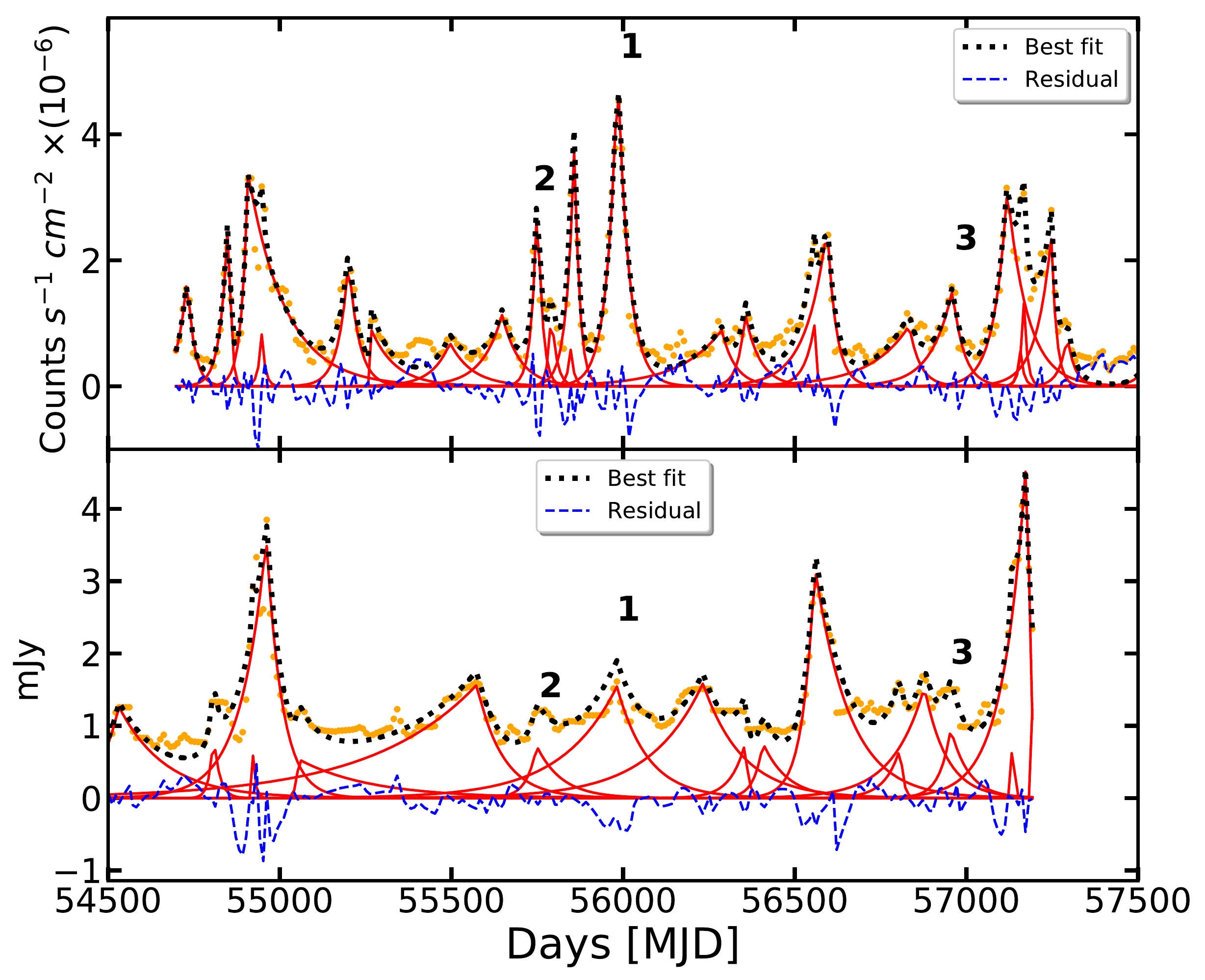}
  \caption{PKS 1510-089}
  \label{fig:sfig1}
\end{subfigure}%
\begin{subfigure}{.5\textwidth}
  \centering
  \includegraphics[width=1.0\linewidth]{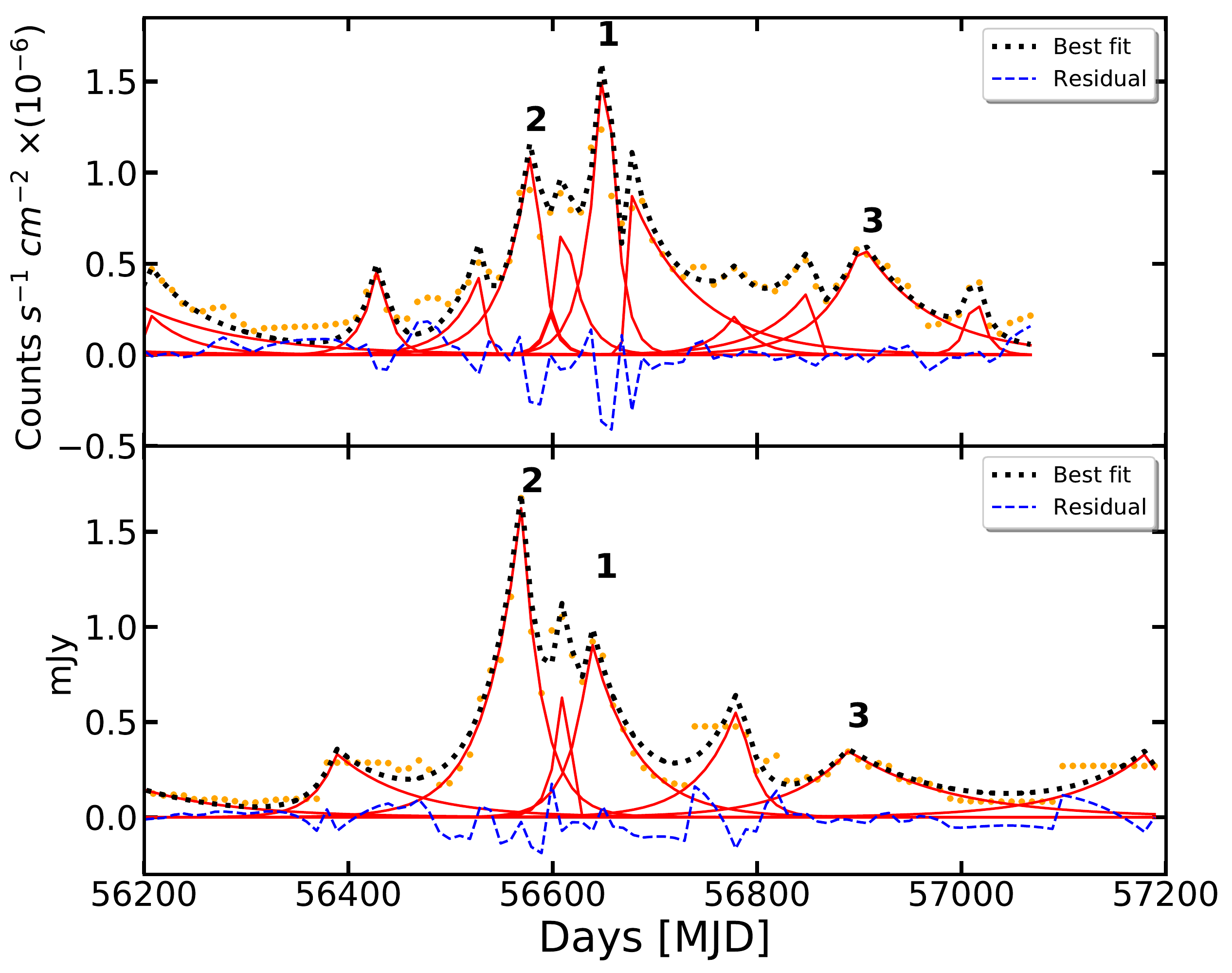}
  \caption{PKS 2326-502}
  \label{fig:sfig2}
\end{subfigure}
 \caption{In the left and right panels we show the simultaneous GeV and optical light curves of PKS 1510-089 and PKS 2326-502, respectively. Yellow filled circles show the observed GeV and $R$-band light curves, smoothed with a Gaussian kernel of standard deviation 10 days, in the upper and lower panels. The red solid lines are the model fitted to individual flares (see Equation \ref{eq:eq1}), the black dotted line is a sum of all the fitted flares and the blue dashed line is the residual after the fit. We perform a similar analysis on all 11 blazars of our sample. We identify flare pairs from the GeV and optical light curves by looking at near simultaneous outbursts at both the bands separated by at most 20 days. Outbursts marked with the same number at both the top and bottom panels are identified flare pairs. Some GeV and optical outbursts may seem simultaneous in this compact figure but they are not identified as flare pairs if their separation is more than 20 days.}
\label{fig:fig1}
\end{figure*}
\section{Data sets}
We use simultaneous $\gamma$-ray ($0.1-300$ GeV) and optical light curves in the $R$-Band from the {\it{Fermi}}-LAT and Yale-SMARTS blazar monitoring programme\footnote{http://www.astro.yale.edu/smarts/glast/home.php}, respectively. Our sample consists of  11 blazars well monitored from 2008-2016. Objects having  $\gamma$-ray light curves with flux values above the monitoring threshold ($10^{-6}$ $cm^{-2}$ $s^{-1}$) throughout the observation period were chosen. The blazars in our sample exhibit multiple contemporaneous outbursts in the GeV and R-band as observed by Fermi-LAT and SMARTS, respectively during the above time interval. The sample is biased towards objects which have shown a significant number ($>$ 10) of prominent flares during the period of monitoring. The final sample was narrowed down based on the presence of the sources in both {\it{Fermi}}-LAT and Yale-SMARTS blazar monitoring programme. 

\subsection{$\gamma$-Ray data}
The GeV data for the target blazars have been obtained from the FSSC website\footnote{https://fermi.gsfc.nasa.gov/ssc/data/access/}. We use the all-sky weekly data files provided by the {\it {Fermi}} team. We select the data from the energy range 0.1$-$300 GeV and the time of observation from 2008 to 2016. We perform unbinned likelihood analysis of the data using \texttt{Fermitools} and with the \texttt{P8R2\_SOURCE\_V6} instrument response function (IRF). We select events with the region of interest (ROI) of \ang{10} around the target blazar with classifications \texttt{evclass} = 128 and \texttt{evtype} = 3.  We select good time interval (GTI) with the filters \text{“DATA\_QUAL==1”} and \text{“LAT\_CONFIG==1”}. To obtain a model of all possible $\gamma$-ray sources within the region of observation we use the 4 year LAT catalog \citep{acero15} given in \textit{gll\_psc\_v16.fit}.  We model the blazar emission with a power-law model keeping the parameters at the catalog value except for the target blazar for which we keep the normalization and index free. For other sources within \ang{3} we keep the normalization a free parameter. We have included Galactic and extragalactic diffuse emission and isotropic background using the templates \textit{gll\_iem\_v06} and \textit{iso\_P8R2\_SOURCE\_V6\_v06} respectively. A source is assumed to be significantly detected in a time bin if the value of the Test Statistic (TS) is greater than 25. We use 1-day time bins for all the blazars in our sample apart from PKS 0208-512 and PKS 1244-255, which did not have significant detection in the 1 day bins. Hence, for those sources we use a 7-day time bin. 
Our complete sample consists of eleven LSP blazars and we present the full sample of objects in Table \ref{tab:table1}. All the objects included in our sample are FSRQs and we have excluded the BL Lacs. The BLR and torus in the BL Lac type objects are weak \citep[e.g see][]{whitting2005, ghisselini2011}. Our analysis, as described in Section 3, is based on the location of the GeV emission region w.r.t. the BLR and torus. Therefore, it may not be directly applicable to BL Lac type sources.

\begin{table*}
\begin{center}

\caption{Table summarizing the properties of the blazars in our sample}
\label{tab:table1}
\begin{tabular}{c|c|c|c|c}
\hline
 Object name & Redshift & TeV detection & Fermi time interval [MJD] & SMARTS time interval [MJD] \\
\hline
3C 279 & 0.536 & Yes & 54683 - 57450 &  54501 - 57404\\
3C 273  & 0.158 & No & 54683 - 57860 & 54537 - 56791 \\
3C 454.3 & 0.859 & No & 54683 - 57064 & 54640 - 57356\\
PKS 0208-512 & 1.003 & No & 54721 - 58509 & 54501 - 57060 \\
PKS 0402+362 & 1.417 & No & 54908 - 57615 & 55838 - 57119 \\
PKS 0454-234 & 1.002 & No & 54690 - 57344 & 55861 - 57145\\
PKS 1244-255  & 0.638 & No & 54693 - 57955 & 55676 - 57115\\
PKS 1424-41 & 1.522 & No & 55001 - 57356 & 55942 - 57200\\
PKS 1510-089 & 0.361 & Yes & 54697 - 57820 & 54501 - 57204\\
PKS 2142-75 & 1.139 & No & 55183 - 58172 & 55297 - 56999\\
PKS 2326-502 & 0.518 & No & 54917 - 57077 &  56109 - 57201 \\
\hline
\end{tabular}
\end{center}
\end{table*}

\subsection{Optical data}
To study the variability of the blazars in our sample we use optical $R$-band light curves from the Yale-SMARTS blazar monitoring programme. The observations were done by the ANDICAM instrument on the SMARTS 1.3 m telescope at CTIO, Chile. All the blazars in the southern sky detected by \textit{Fermi}-LAT were followed up by the SMARTS programme. The observing cadence of a source depended on its $\gamma$-ray brightness and hence the sampling is not uniform. The details of data acquisition and data analyses are given in \citet{bonning2012}. 

\section{Method}
We first analyze the light curves of the 11 blazars of our sample and identify the `flare pairs.' We study the distribution of energy dissipation in the optical and GeV flares. We have developed a theoretical jet emission model from which we attempt to constrain the location of the emission region by studying the ratio of energy emitted at the GeV and optical wavebands. In the following subsections we describe the method to analyze the light curves and identify flare pairs. We also discuss our theoretical model of blazar jet emission.

\subsection{Analysis of the observed light curves}
To fit the individual light curves we follow the prescription of \citet{valtoja99,chatterjee12}. The light curves are smoothed with a Gaussian kernel of standard deviation 10 days. This is done to remove any short timescale variability features and we focus on timescales related to the shock crossing timescales.  After this we identify the highest peak in the light curve and fit it with a double exponential function with four fitting parameters (see Equation \ref{eq:eq1}): rise timescale (${\tau}_{rise}$), fall timescale (${\tau}_{fall}$), amplitude of flare ($A_{0}$) and epoch of flare ($t_{0}$).  Then we subtract the fitted flare from the light curve and apply the same procedure on the residual light curve. We continue this process till the highest flare amplitude in the residual light curve is 10\% of the original light curve’s largest flare.
\begin{equation} \label{eq:eq1}
    f(t)=
\begin{cases}
A_{0}e^{\frac{t-t_{0}}{{\tau}_{rise}}} & \text{if }   t<t_{0} \\
A_{0}e^{\frac{t_{0}-t}{{\tau}_{fall}}} & \text{if }   t>t_{0} \\
\end{cases}
\end{equation}
In Figure \ref{fig:fig1}, we show the GeV and R-Band light curves of the blazars PKS 1510-089 and PKS 2326-502, along with the individual fitted flares (solid red lines), the sum of all the fitted flares (black dotted line) and the residual (blue dashed line). In a similar fashion we have decomposed the light curves of the 11 blazars in our sample and identified 305 GeV and 196 optical flares. We define flare pairs as simultaneous or near simultaneous outbursts in different electromagnetic bands. We identify two flares in the GeV and optical bands as a flare pair if their best-fit $t_{0}$ values (time of peak flux of the flare) are separated by less than 20 days. 
Using this criteria we identified 70 flare pairs. The identified flare pairs for the two systems discussed here have been marked in Figures \ref{fig:sfig1} and \ref{fig:sfig2}. Given our strict criteria for selecting flare pairs, it is possible that we may exclude some pairs but it may be safely assumed that our identified flare pairs are simultaneous and hence caused by the same event in the jet. 

We obtain the energy dissipated by an outburst by integrating the area under the fitted flare function. The GeV flares were converted from counts to energy units using the conversion factor given by Equation \ref{eq:eq2}, in which ${\nu}_{0}$ and ${\nu}_{1}$ represent the lowest and highest frequency of the relevant energy band, respectively. The formula is obtained assuming a power-law spectral model for the blazars. $\alpha$ is given by $1-\Gamma$, where $\Gamma$ is the average photon index of the blazar. The values of $\Gamma$ have been obtained from the 4 year LAT catalogue \citet{acero15}.
\begin{equation} \label{eq:eq2}
    ECF=h{{\nu}_{0}}\frac{({\nu}_{1}/{\nu}_{0})^{\alpha+1}-1}{({\nu}_{1}/{\nu}_{0})^{\alpha}-1}\frac{\alpha}{\alpha+1}
\end{equation}

\begin{figure}
\centering
         \includegraphics[width=\linewidth]{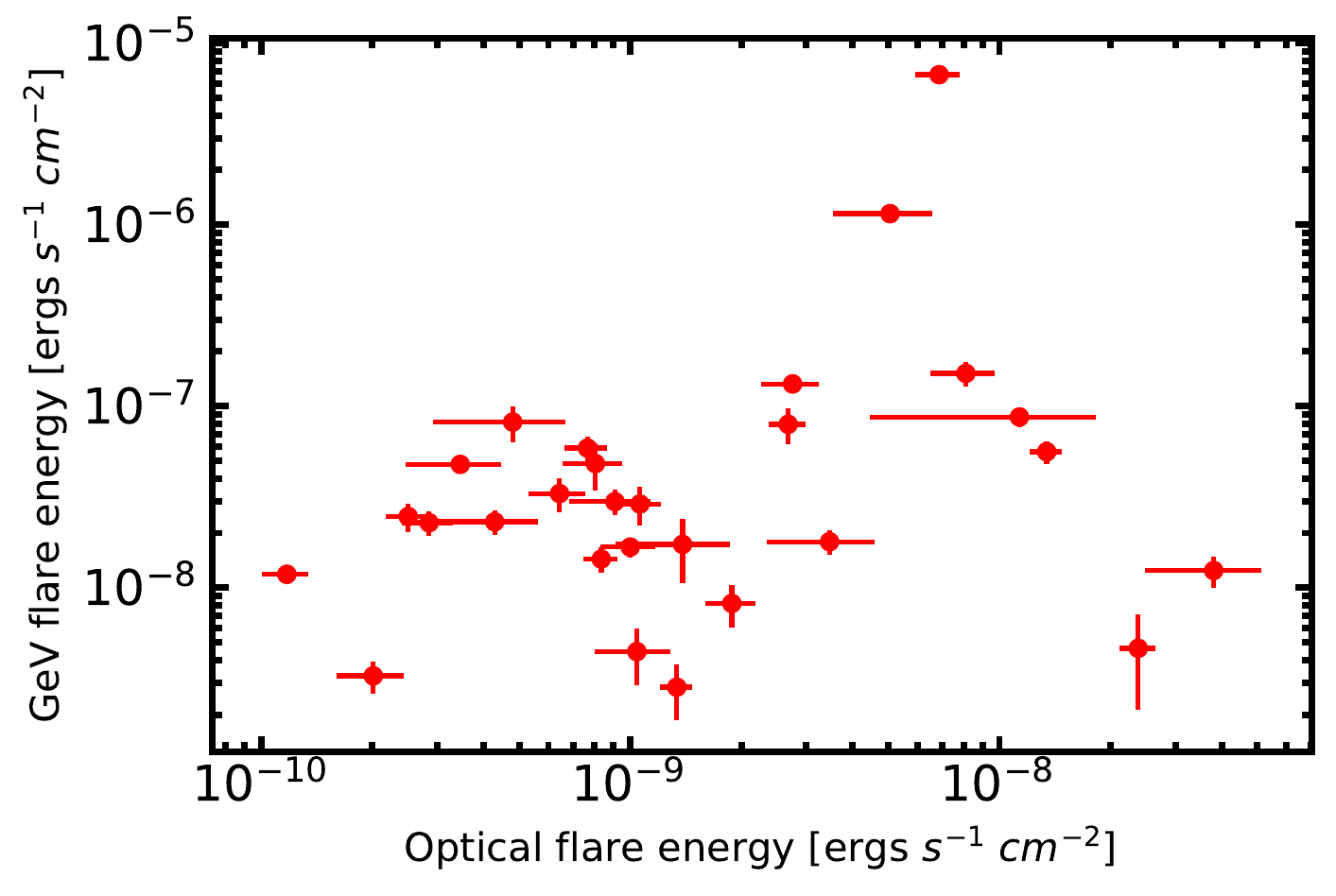} 
\caption{The red filled circles with error bars denote the energy dissipated in GeV and optical bands for 25 flare pairs in the 11 blazars in our sample. In those flare pairs, the uncertainty of the rise and fall timescales for both the optical and GeV flares are less than 50\%. This sample includes only FSRQ objects.}
\label{fig:fig2}
\end{figure}

In Figure \ref{fig:fig2}, we can see that the the GeV outbursts are more energetic compared to the optical flares. The main sources of error on the computed value of the dissipated energy of the flares are the rise and fall timescales of the flares which determine the integration limits. In order to avoid flare pairs for which the parameters have not been determined precisely, we select a sample of flare pairs, in which the uncertainties of the rise and fall timescales for both GeV and R-band flares are less than $50\%$. Our final sample consists of 25 flare pairs spread over 11 LSP blazars and 8 years of data. 

\subsection{Theoretical jet emission model}
In our computer code, we use a simplified model of the emission region. The entire emission region is assumed to be cylindrical. It is divided into multiple cells with each cell having its own magnetic field and population of electrons. The evolution of the energy distribution of electrons in the individual cells are assumed to be independent of each other. We only consider bulk motion of the plasma along the axial direction of the jet and any direction normal to the axis is not of significance to our simulation. The magnetic field in each cell goes as $B \sim r^{-1}$ where $r$ is the distance from the central engine. We assume a magnetic field strength at the base and tip of the emission region, denoted by $B_{i}$ and $B_{f}$, respectively. 
\begin{figure}
    \centering      
  \includegraphics[width=\columnwidth]{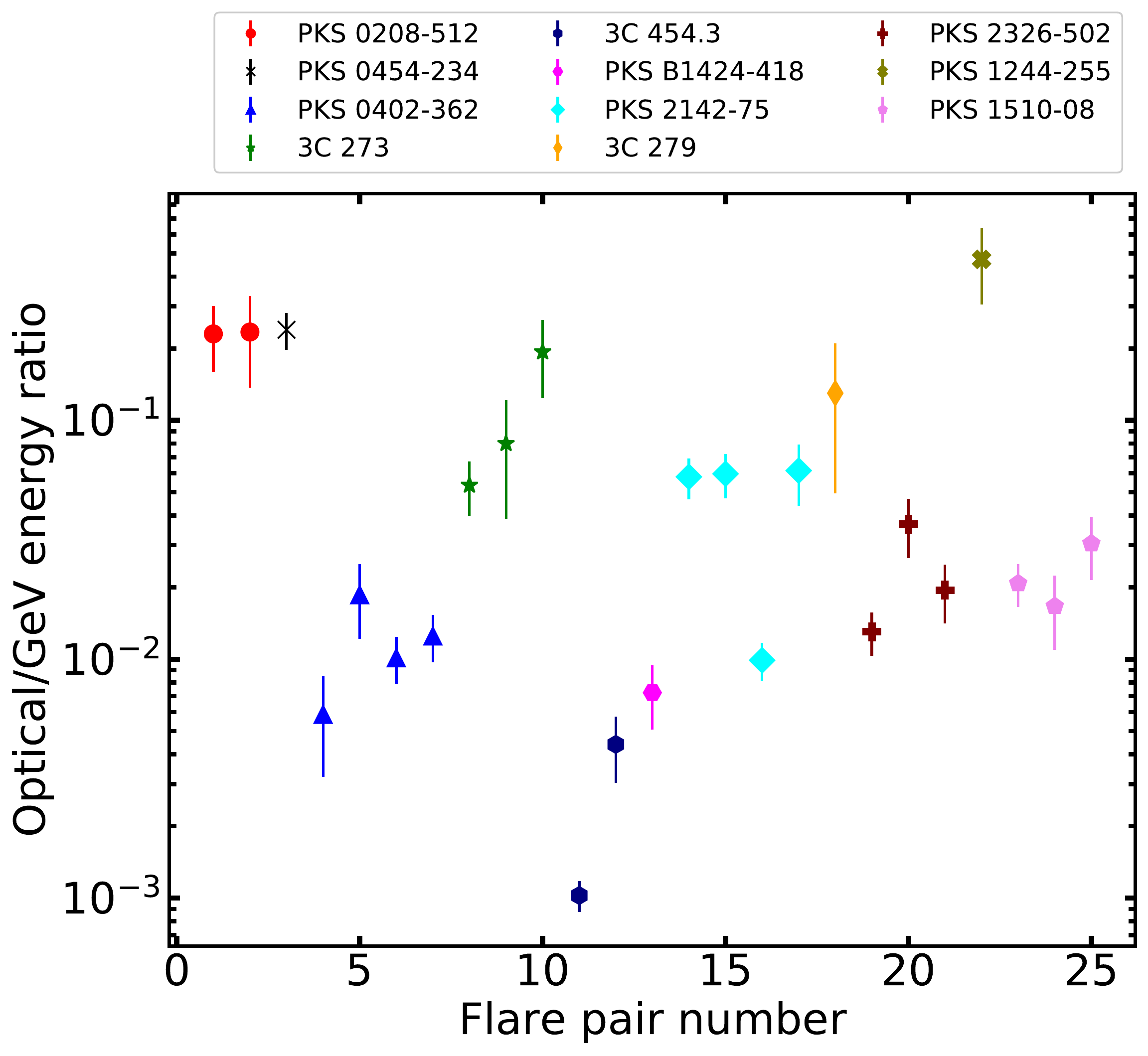} 	 
 \caption{Points denote the ratio of energy dissipated at the $R-$band to that at the GeV band for the flare pairs identified in our sample. Points of same color and type belong to the same blazar. We assign a serial number to each flare pair in our sample for the sake of identification. The serial number is plotted in the X-axis. }
\label{fig:fig3}
\end{figure}

It is generally assumed that the emitting particles in the jet are accelerated to relativistic energies by their interaction with shock waves moving down the jet, which results in the high amplitude long-term ($\sim$weeks to months) outbursts in blazars \citep[e.g.,][]{lazzati99,marscher1985}, as studied here. In our model, we consider the shock to be a front that moves through the cells of the emission region and instantaneously energize the electron population within that region to a power-law energy distribution given by $N(\gamma)=N_{0}{\gamma}^{-s}$, where $s = 2.5$ and $\gamma$ is the electron Lorentz factor. The shock front is assumed to start at the base of the emission region and traverse the length of the entire region at a speed close to that of light.

The electron energies range from ${\gamma}_{min}$ to ${\gamma}_{max}$. The energized electrons cool by synchrotron radiation and inverse-Compton scattering. We include both synchrotron self-Compton (SSC) and external Compton (EC) processes in our model. The synchrotron photons from all the other cells are considered keeping in mind the light travel time delays and the inverse squared dependence on distance while calculating the seed photon field for SSC. We assume the seed photon field for EC to be generated from the broad line region (BLR) and dusty torus. The seed photon field densities are given by \citet{hayashida12}
\begin{equation} \label{eq:eq3}
    u^{'}_{BLR}(r)=\frac{{\epsilon}_{BLR}{\Gamma}^{2}_{jet}L_{D}}{3{\pi}r^{2}_{BLR}c[1+(r/r_{BLR})^{{\beta}_{BLR}}]}
\end{equation}

\begin{equation} \label{eq:eq4}
    u^{'}_{torus}(r)=\frac{{\epsilon}_{torus}{\Gamma}^{2}_{jet}L_{D}}{3{\pi}r^{2}_{torus}c[1+(r/r_{torus})^{{\beta}_{torus}}]},
\end{equation}
where $r$ is the distance of the cell from the central engine, ${\epsilon}_{BLR} = 0.1$ and ${\epsilon}_{torus} = 0.01$ represent the fraction of accretion disc luminosity that is reprocessed by the BLR and torus, respectively, and re-emitted toward the jet. The BLR and torus photons are injected into the photon field as monochromatic emission in the ultraviolet and infrared frequencies respectively. Following the prescription of \citet{hayashida12}, ${\beta}_{BLR}$ = 3 and ${\beta}_{torus}$ = 4. We use $r_{BLR}~=~0.1(L_{D,46})^{1/2}pc$ and $r_{torus}~=~2.5(L_{D,46})^{1/2}pc$ following \citet[][]{hayashida12}, where $L_{D,46}$ is the disk luminosity in the units of $10^{46}$ ergs\,s$^{-1}$, and ${\Gamma}_{jet}$ = 10.

\section{Results} \label{sec:section4}
In Figure \ref{fig:fig3}, we show the ratio of the energy emitted at the $R$-band to that at the GeV frequencies for the flare pairs detected in our sample. 
We see that in most cases the said ratio is much less than unity. We take the cases of two particular blazars PKS-1510-089 and PKS 2326-502 with three identified flare pairs each and constrain the location of the emission region by comparing the observed ratios with those produced by the theoretical model. We choose these blazars because we identified three or more prominent flares in these two blazars with smaller relative uncertainties of the best-fit parameters of the flares, which can lead to well-determined values of the energy dissipation ratio, and their SEDs are well constrained. We estimate the effect on the measured energy ratios due to the residuals in the fitted light curves. For the case of PKS 1510-089, we find that the rms of the  residuals in the GeV light curve is 7\% of the amplitude of the largest flare and the corresponding rms in the optical light curves is 10\%. Propagating these uncertainties into the energy ratio, we find the uncertainty to be 13\% which is smaller than the estimated error bars. We conclude similarly for PKS 2326-502. In the simulation we keep certain parameters, e.g., ${\gamma}_{max}$, ${\Gamma}_{jet}$ and magnetic field fixed to appropriate values as indicated by their observed SED found in the literature, and vary the location of the emission region to identify which location produces an optical/GeV energy dissipation ratio that matches with the above values as shown in Figure \ref{fig:fig3}. The parameters we have used for simulating the light curves of these blazars have been summarized in Table 2. A follow-up paper, in which the location of the emission region of a larger sample of blazar flares in our sample will be investigated, is in preparation.

\begin{figure*}[htp!]
        \centering
         \resizebox{15cm}{!}{\includegraphics[width=\linewidth]{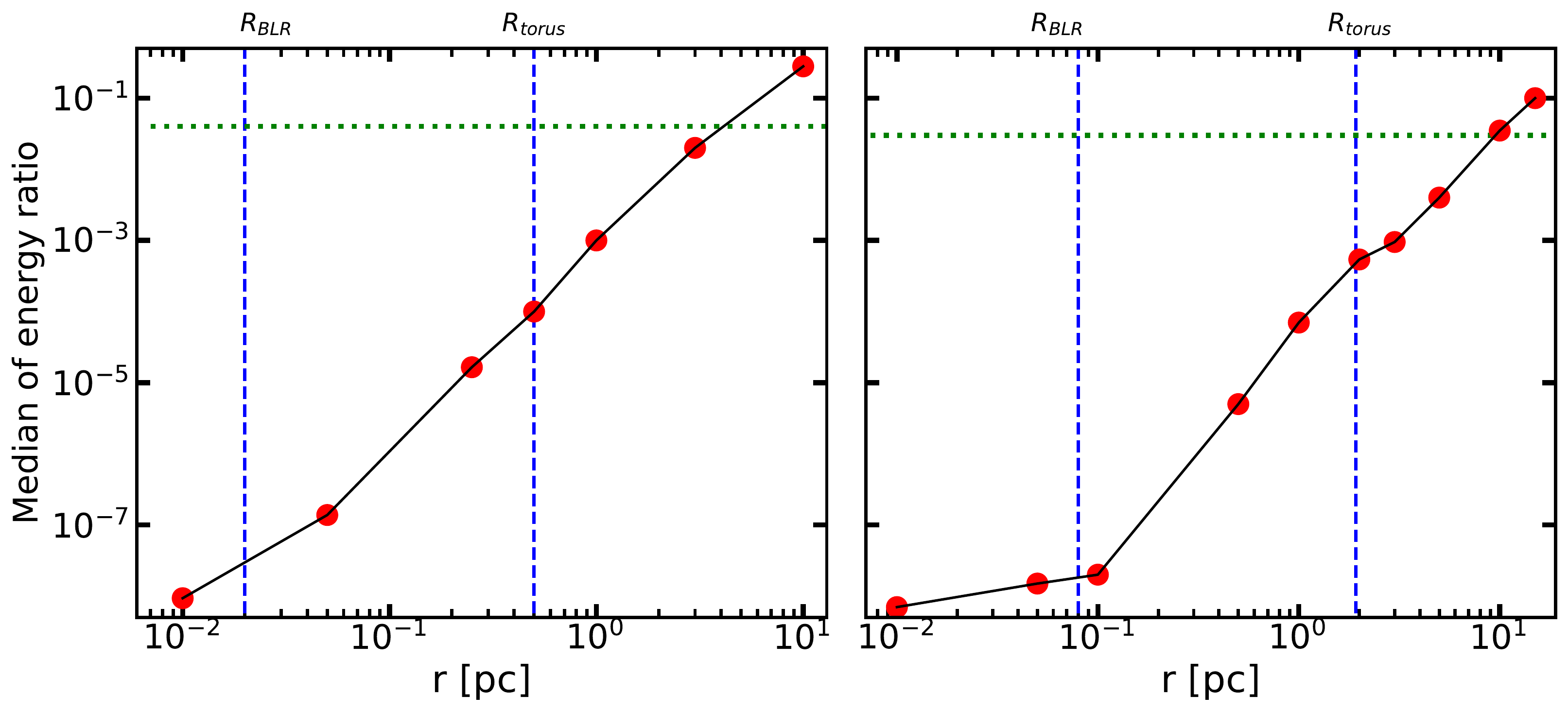} }
   
          
 \caption{In the left panel, the black solid line denotes the median of the optical to GeV band energy dissipation ratio of the flare pairs in the simulated light curves of PKS 2326-502. The light curves were simulated using ${\gamma}_{max}=10000$, ${\Gamma}_{jet}=10$ and $L_{disk}=3 \times 10^{44} erg s^{-1}$. The blue dashed lines represent the location of the BLR (0.02 pc) and torus (0.5 pc), respectively and the green dotted line shows the median value (0.04) of the energy dissipation ratio we find from the observed light curves of PKS 2326-502. In the right panel, the black solid line denotes the median of the optical to GeV band energy dissipation ratio of the flare pairs in the simulated light curves of PKS 1510-089. The light curves were simulated using ${\gamma}_{max}=3000$, ${\Gamma}_{jet}=10$ and $L_{disk}=5.9 \times 10^{45} erg s^{-1}$. The blue dashed lines represent the location of the BLR (0.08 pc) and torus (1.92 pc), respectively and the green dotted line shows the median value (0.03) of the energy dissipation ratio we find from the observed light curves of PKS 1510-089.}  \label{fig:fig4}
\end{figure*}
\subsection{PKS 1510-089}
In Figure \ref{fig:fig3}, we can see the optical/GeV energy ratio for PKS 1510-089 for all the three flare pairs detected in the light curves of this blazar denoted by the filled violet pentagons. The median value of the ratio for these three flare pairs is $\sim 0.03$. We have used the detailed broadband SED of PKS 1510-089 from \citet{Abdo_sed}. The synchrotron peak is between $10^{12}$ to $10^{13}$ Hz. PKS 1510-089 is an LSP blazar and from our model we find that the observed synchrotron peak of this blazar may be reproduced using ${\gamma}_{max}=3000$. We have fixed the value of ${\Gamma}_{jet}$ = 10 for the purpose of this work. The magnetic field is held at 0.5 Gauss at the base of the emission region and 0.3 Gauss at the tip following the results of \citet{majumdar_19}. The values of the above parameters are consistent with its observed SED. 
Using $L_{disk}=5.9 \times 10^{45} erg\,s^{-1}$ \citep[][]{castignanai17} we calculate $r_{BLR}$ = 0.08 pc and $r_{torus}$ = 1.92 pc. 
In Figure \ref{fig:fig4} right panel, we show the median of the ratio of energy dissipated in the simulated light curves of PKS 1510-089 as a function of the distance of the emission region from the central black hole. The blue dashed lines show the location of the BLR and IR emitting regions (the torus) and the green dotted line shows the median of the observed optical/GeV energy ratio distribution for the PKS 1510-089 flare pairs. We can see that the location of the emission region should be around 10 pc, far away from the BLR and slightly beyond the molecular torus, to match with the median of the observed values of optical/GeV energy ratio. The reason for this is that the GeV flares we probe using \textit{Fermi}-LAT are predominantly produced due to the EC process with seed photons coming mostly from the BLR and partly from the molecular torus as can be seen from the results in \citet{namrata_19} and discussions therein. Thus, if the emission region is close to the BLR ($\sim$0.1 pc) the BLR seed photon density is very high and it produces strong GeV flares resulting in very small optical/GeV energy ratio. As we go further out from the BLR and torus length scales, the seed photon field density falls rapidly and the GeV flares get weaker, so from our simulations we conclude that 10 pc is roughly the distance where the GeV emission region can occur for PKS 1510-089 based on the observed data. In Section 5, we discuss related results for PKS 1510-089 obtained by different authors using various arguments. 

\subsection{PKS 2326-502}
The optical/GeV energy ratios observed for the blazar PKS 2326-502 are plotted in Figure \ref{fig:fig3} with maroon crosses. The synchrotron peak in the SED of PKS 2326-502 is $\sim 10^{14}$ Hz \citep{dutka17}, which may be reproduced using ${\gamma}_{max}=10000$ in our model. Similar to the previous case, we fix ${\Gamma}_{jet}$=10 and the magnetic field at the base of the emission region at 0.5 G and at the tip of the jet at 0.3 G. We use $L_{disk} = 3 \times 10^{44} erg s^{-1}$ \citep[][]{dutka17} and calculate $r_{BLR} =$ 0.02 pc and $r_{torus} =$ 0.5 pc. A summary of the parameters used to simulate the light curves for this system are given in Table \ref{tab:table2}. In Figure \ref{fig:fig4} left panel, we show the simulated optical/GeV flare energy ratio as a function of distance from the central black hole. The green dotted line shows the median of the observed optical/GeV flare energy ratio for this blazar ($\sim 0.02$). Similar to the case of PKS 1510-089, we find that the location of the emission region lies much beyond the BLR, and slightly beyond the molecular torus at roughly 4 pc.

\begin{table}
\caption{Table summarizing parameters used for simulating the light curves of PKS 1510-089 and PKS 2326-502.}
\label{tab:table2}
\begin{tabular}{c|c|c}
\hline
 Parameter & PKS 1510-089 & PKS 2326-502 \\
\hline
${\Gamma}_{jet}$ & 10 & 10\\
redshift & 0.361 & 0.518\\
SED reference & \citet[][]{Abdo_sed} & \citet[][]{dutka17}\\
${\gamma}_{max}$ & 3000 & 10000\\
${\gamma}_{min}$ & 1 & 1\\
Power law index (s) & -2.5 &-2.5\\
$B_{i}$ & 0.5G & 0.5G\\
$B_{f}$ & 0.3G & 0.3G\\
$L_{disk}$ & 5.9$\times 10^{45}$ ergs/s \footnote{\citet[][]{castignanai17}} & 3 $\times 10^{44}$ ergs/s \footnote{\citet{dutka17}}\\
$R_{BLR}$ & 0.08pc & 0.02pc\\
$R_{IR}$ &1.92pc & 0.5pc\\
${\eta}_{BLR}$ & 0.1 & 0.1\\
${\eta}_{IR}$ & 0.01 & 0.01\\
${\beta}_{BLR}$ & 3 & 3\\
${\beta}_{IR}$ & 4 & 4\\
\hline
\end{tabular}
\end{table}

\section{Discussion} \label{sec:sec5}
Locating the GeV emission region in the jets of blazars is an open problem and an active area of research. Here, we have discussed the possibility to constrain the location of the GeV emission region by probing the ratio of energy emitted in the optical/GeV flare pairs. We assume that GeV emission observed in the blazars in our sample is due to EC and the contribution of the SSC process in it is negligible. The blazars in our sample are known LSP blazars and from simulated and observed SEDs of such blazars \citep[e.g.][]{namrata_19,Abdo_sed,Meyer_2012} we can see that the SSC contribution is dominant in the X-ray regime and drops off at the GeV band. We find that the energy ratio, which is essentially a ratio of synchrotron to EC emission, strongly depends on the location of the emission region w.r.t. the source of seed photons and hence the distance of the former from the central engine.  

We apply this approach to two individual blazars, namely, PKS 1510-089 and PKS 2326-502, to constrain the location of GeV emission in them. The models that we use for our simulations depend on various parameters of the blazar such as ${\gamma}_{max}$, $L_{disk}$, the magnetic field strength ($B$) and location of the BLR and torus. We use values of ${\gamma}_{max}$ and $L_{disk}$ as indicated by the modeling of observed SEDs published in the literature. We derive $R_{BLR}$ and $R_{torus}$ from scaling relations discussed in \citet[][]{hayashida12}. Those estimates have uncertainties, but the expression of seed photon energy density from the BLR and the torus in \citet[][]{hayashida12} implies that if the emission region is within the BLR (or torus), the seed photon field density is essentially constant and outside this region, the density falls off rapidly. Hence, the absolute locations of the BLR and torus do not have an impact on our conclusion that the emission region lies beyond the BLR. The strength of the magnetic field does affect the optical/GeV energy ratio because it determines the total energy dissipated in the optical flares which originate due to the synchrotron process. From our model we find that the effect of changing the magnetic field strength at the base of the jet on the energy dissipation scales as $B^{2}$. We ran simulations where we varied the strength of the magnetic field from 0.1 G to 5 G. We found that the energy ratio varies by roughly three orders of magnitude, which is significant but much smaller than the effect of changing the location of the emission region. Therefore, for the purpose of this work we fix the value of the magnetic field strength. 


Constraining the location of GeV emission in blazars has been an important problem in the field and several authors have discussed this problem with different arguments for large samples or individual blazars including PKS 1510-089. Using SED modeling during a VHE $\gamma$-ray flare \citet{Ahnen2017} concluded that the emission region in PKS 1510-089 may be the VLBA core, which is located much beyond the torus at $\sim$10 pc \citep{pushkarev2012,aleksi14}. In that case, the seed photons are supplied by the sheath of the jet \citep{aleksi14,macdonald15}. Based on the observed variability timescale of the VHE $\gamma$-rays and assuming suitable geometric parameters of the jet found in the literature, they suggested that the emission region may be located closer, e.g., at $\sim$2.7 pc. On the other hand, using SED modeling, \citet{roy2021} found that the GeV emission region in PKS 1510-089 is closer to the outer edge of the BLR and away from the torus. The location of the emission region has important implications about the particle acceleration mechanisms and radiation processes. \citet{Dotson_2012} suggest that, if the emission region is located near the BLR the cooling timescales for short timescale (few days) flares would be in the Klein-Nishina regime and would be achromatic whereas if it is near the molecular torus it is expected to be in the Thomson regime and we expect faster variability at higher energies. \citet[][]{Dotson_2015, Saito_2013} have studied the decay timescales of short timescale GeV flares of PKS 1510-089 and have confirmed the location of the GeV emission region to be beyond the BLR. Other authors also conclude that the emission region is beyond the BLR based on broadband spectral features \citep[][]{castignanai17} and the absence of $\gamma$-$\gamma$ absorption signature in the GeV spectra of more than hundred Fermi blazars \citep[][]{costamate18}, including PKS 1510-089. Although, a large body of work suggests that GeV emission can originate predominantly a few pc down the jet from the central engine, it poses challenges for particle acceleration mechanism and transportation before the energized particles have time to cool. Some authors have suggested that GeV emission may be generated at multiple active emission zones within the jet. For example, \citet[][]{brown13} suggest multiple emission zones for PKS 1510-089. Using variability timescales, $\gamma$-ray spectral cut-off and the energy dependence of cooling timescales, \citet{acharyya2021} have concluded that GeV emission in several FSRQs, including PKS 1510-089, originate at multiple regions along the jet, within and beyond the BLR. Similar conclusion has been drawn by \citet{prince2019} using SED modeling, and \citet{marscher2010} by analyzing simultaneous multi-wavelength monitoring of GeV flux, optical flux and polarization and pc-scale dynamics with the VLBA. 

We conclude that for two blazars in our sample, PKS 1510-089 and PKS 2326-502, the emission region is located beyond the BLR by comparing the observed energy ratio to the results from a theoretical model we have developed. In addition, our results show that the jet is able to generate even very bright GeV outbursts paired with simultaneous flaring in the optical at locations much beyond the BLR. This provides strong constraints on the models of energy injection, dissipation, and transport in the blazar jet.

Using the ratios of energy dissipated in flare pairs is a new method to constrain the location of GeV emission region for LSP blazars, which is complementary to other techniques used in the literature. Comparison of our results with the location obtained by other methods will be useful in making the methods more precise. For example,  \citet{Rani2018}, using simultaneous monitoring of multi-wavelength flux, polarization and pc-scale dynamics with VLBI, concluded that one out of five multi-wavelength flares (designated by them as ``flare 5'') of 3C 279 during an active epoch in 2013-14 originated upstream of the VLBI core, i.e., closer to the central engine while the other flares originated farther out at the core. We made an estimate of the optical/GeV energy ratio of flare 5 and flare 3 of \citet{Rani2018}. We find that for the two adjacent flares during the epoch of flare 5 the ratios are significantly smaller than that for flare 3. It indicates the location of the former may be closer to the central engine than the latter, which is consistent with the findings of \citet{Rani2018}. However, a more detailed analysis is required to make more precise estimate of the location. A set of analyses similar to what we have carried out in this paper but for a larger sample of flare pairs in the blazars in our sample, particularly for which estimates of the location are available from other methods, is in preparation.  

\section*{ACKNOWLEDGEMENTS}
SB thanks Sagnick Mukherjee for helpful discussion. SB is grateful to Jean-Michel Désert and the University of Amsterdam for support.
RC thanks Presidency University for support under the Faculty Research and Professional Development (FRPDF) Grant, ISRO for support under the AstroSat archival data utilization program, and IUCAA for their hospitality and usage of their facilities during his stay at different times as part of the university associateship program. RC acknowledges financial support from BRNS through a project grant (sanction no: 57/14/10/2019-BRNS) and thanks the project coordinator Pratik Majumdar for support regarding the BRNS project. 
\section*{Data Availability}
The data utilized in this work have been taken from:-
\begin{enumerate}
    \item Fermi 4FGL-DR2 and 4LAC catalogues.\\ Weblink: \url{https://fermi.gsfc.nasa.gov/ssc/data/access/lat/10yr_catalog/}
    
    \item  SMARTS optical light curves\\ Weblink: \url{www.astro.yale.edu/smarts/glast/home.php}
    
\end{enumerate}

\bibliography{blazar.bib}

\begin{thebibliography}{}
\makeatletter
\relax
\def\mn@urlcharsother{\let\do\@makeother \do\$\do\&\do\#\do\^\do\_\do\%\do\~}
\def\mn@doi{\begingroup\mn@urlcharsother \@ifnextchar [ {\mn@doi@}
  {\mn@doi@[]}}
\def\mn@doi@[#1]#2{\def\@tempa{#1}\ifx\@tempa\@empty \href
  {http://dx.doi.org/#2} {doi:#2}\else \href {http://dx.doi.org/#2} {#1}\fi
  \endgroup}
\def\mn@eprint#1#2{\mn@eprint@#1:#2::\@nil}
\def\mn@eprint@arXiv#1{\href {http://arxiv.org/abs/#1} {{\tt arXiv:#1}}}
\def\mn@eprint@dblp#1{\href {http://dblp.uni-trier.de/rec/bibtex/#1.xml}
  {dblp:#1}}
\def\mn@eprint@#1:#2:#3:#4\@nil{\def\@tempa {#1}\def\@tempb {#2}\def\@tempc
  {#3}\ifx \@tempc \@empty \let \@tempc \@tempb \let \@tempb \@tempa \fi \ifx
  \@tempb \@empty \def\@tempb {arXiv}\fi \@ifundefined
  {mn@eprint@\@tempb}{\@tempb:\@tempc}{\expandafter \expandafter \csname
  mn@eprint@\@tempb\endcsname \expandafter{\@tempc}}}

\bibitem[\protect\citeauthoryear{{Abdo} et~al.,}{{Abdo}
  et~al.}{2010}]{Abdo_sed}
{Abdo} A.~A.,  et~al., 2010, \mn@doi [\apj] {10.1088/0004-637X/716/1/30}, \href
  {https://ui.adsabs.harvard.edu/abs/2010ApJ...716...30A} {716, 30}

\bibitem[\protect\citeauthoryear{{Abdollahi} et~al.,}{{Abdollahi}
  et~al.}{2020}]{abdolahi20}
{Abdollahi} S.,  et~al., 2020, \mn@doi [\apjs] {10.3847/1538-4365/ab6bcb},
  \href {https://ui.adsabs.harvard.edu/abs/2020ApJS..247...33A} {247, 33}

\bibitem[\protect\citeauthoryear{{Acero} et~al.,}{{Acero}
  et~al.}{2015}]{acero15}
{Acero} F.,  et~al., 2015, \mn@doi [\apjs] {10.1088/0067-0049/218/2/23}, \href
  {https://ui.adsabs.harvard.edu/abs/2015ApJS..218...23A} {218, 23}

\bibitem[\protect\citeauthoryear{{Acharyya}, {Chadwick}  \& {Brown}}{{Acharyya}
  et~al.}{2021}]{acharyya2021}
{Acharyya} A.,  {Chadwick} P.~M.,   {Brown} A.~M.,  2021, \mn@doi [\mnras]
  {10.1093/mnras/staa3483}, \href
  {https://ui.adsabs.harvard.edu/abs/2021MNRAS.500.5297A} {500, 5297}

\bibitem[\protect\citeauthoryear{{Agudo} et~al.,}{{Agudo}
  et~al.}{2011}]{Agudo_2011}
{Agudo} I.,  et~al., 2011, \mn@doi [\apjl] {10.1088/2041-8205/726/1/L13}, \href
  {https://ui.adsabs.harvard.edu/abs/2011ApJ...726L..13A} {726, L13}

\bibitem[\protect\citeauthoryear{{Ahnen} et~al.,}{{Ahnen}
  et~al.}{2017}]{Ahnen2017}
{Ahnen} M.~L.,  et~al., 2017, \mn@doi [\aap] {10.1051/0004-6361/201629960},
  \href {https://ui.adsabs.harvard.edu/abs/2017A&A...603A..29A} {603, A29}

\bibitem[\protect\citeauthoryear{{Aleksi{\'c}} et~al.,}{{Aleksi{\'c}}
  et~al.}{2014}]{aleksi14}
{Aleksi{\'c}} J.,  et~al., 2014, \mn@doi [\aap] {10.1051/0004-6361/201423484},
  \href {https://ui.adsabs.harvard.edu/abs/2014A&A...569A..46A} {569, A46}

\bibitem[\protect\citeauthoryear{{Arbeiter}, {Pohl}  \&
  {Schlickeiser}}{{Arbeiter} et~al.}{2005}]{Arbeiter_2005}
{Arbeiter} C.,  {Pohl} M.,   {Schlickeiser} R.,  2005, \mn@doi [\apj]
  {10.1086/430118}, \href
  {https://ui.adsabs.harvard.edu/abs/2005ApJ...627...62A} {627, 62}

\bibitem[\protect\citeauthoryear{{Arsioli} \& {Chang}}{{Arsioli} \&
  {Chang}}{2018}]{arsioli2018}
{Arsioli} B.,  {Chang} Y.~L.,  2018, \mn@doi [\aap]
  {10.1051/0004-6361/201833005}, \href
  {https://ui.adsabs.harvard.edu/abs/2018A&A...616A..63A} {616, A63}

\bibitem[\protect\citeauthoryear{{Blandford} \& {Levinson}}{{Blandford} \&
  {Levinson}}{1995}]{blanford1995}
{Blandford} R.~D.,  {Levinson} A.,  1995, \mn@doi [\apj] {10.1086/175338},
  \href {https://ui.adsabs.harvard.edu/abs/1995ApJ...441...79B} {441, 79}

\bibitem[\protect\citeauthoryear{{B{\l}a{\.z}ejowski}, {Sikora}, {Moderski}  \&
  {Madejski}}{{B{\l}a{\.z}ejowski} et~al.}{2000}]{B_a_ejowski_2000}
{B{\l}a{\.z}ejowski} M.,  {Sikora} M.,  {Moderski} R.,   {Madejski} G.~M.,
  2000, \mn@doi [\apj] {10.1086/317791}, \href
  {https://ui.adsabs.harvard.edu/abs/2000ApJ...545..107B} {545, 107}

\bibitem[\protect\citeauthoryear{{Bonning} et~al.,}{{Bonning}
  et~al.}{2012}]{bonning2012}
{Bonning} E.,  et~al., 2012, \mn@doi [\apj] {10.1088/0004-637X/756/1/13}, \href
  {https://ui.adsabs.harvard.edu/abs/2012ApJ...756...13B} {756, 13}

\bibitem[\protect\citeauthoryear{{B{\"o}ttcher} \& {Els}}{{B{\"o}ttcher} \&
  {Els}}{2016}]{bottcher2016}
{B{\"o}ttcher} M.,  {Els} P.,  2016, \mn@doi [\apj]
  {10.3847/0004-637X/821/2/102}, \href
  {https://ui.adsabs.harvard.edu/abs/2016ApJ...821..102B} {821, 102}

\bibitem[\protect\citeauthoryear{{Bregman}, {Lebofsky}, {Aller}, {Rieke},
  {Aller}, {Hodge}, {Glassgold}  \& {Huggins}}{{Bregman}
  et~al.}{1981}]{bregman_1981}
{Bregman} J.~N.,  {Lebofsky} M.~J.,  {Aller} M.~F.,  {Rieke} G.~H.,  {Aller}
  H.~D.,  {Hodge} P.~E.,  {Glassgold} A.~E.,   {Huggins} P.~J.,  1981, \mn@doi
  [\nat] {10.1038/293714a0}, \href
  {https://ui.adsabs.harvard.edu/abs/1981Natur.293..714B} {293, 714}

\bibitem[\protect\citeauthoryear{{Brown}}{{Brown}}{2013}]{brown13}
{Brown} A.~M.,  2013, \mn@doi [\mnras] {10.1093/mnras/stt218}, \href
  {https://ui.adsabs.harvard.edu/abs/2013MNRAS.431..824B} {431, 824}

\bibitem[\protect\citeauthoryear{Cao \& Wang}{Cao \& Wang}{2013}]{cao2013}
Cao G.,  Wang J.-C.,  2013, \mn@doi [Monthly Notices of the Royal Astronomical
  Society] {10.1093/mnras/stt1723}, 436, 2170

\bibitem[\protect\citeauthoryear{{Castignani} et~al.,}{{Castignani}
  et~al.}{2017}]{castignanai17}
{Castignani} G.,  et~al., 2017, \mn@doi [\aap] {10.1051/0004-6361/201629775},
  \href {https://ui.adsabs.harvard.edu/abs/2017A&A...601A..30C} {601, A30}

\bibitem[\protect\citeauthoryear{{Chatterjee} et~al.,}{{Chatterjee}
  et~al.}{2012}]{chatterjee12}
{Chatterjee} R.,  et~al., 2012, \mn@doi [\apj] {10.1088/0004-637X/749/2/191},
  \href {https://ui.adsabs.harvard.edu/abs/2012ApJ...749..191C} {749, 191}

\bibitem[\protect\citeauthoryear{{Chavushyan}, {Pati{\~n}o-{\'A}lvarez},
  {Amaya-Almaz{\'a}n}  \& {Carrasco}}{{Chavushyan} et~al.}{2020}]{chavushyan20}
{Chavushyan} V.,  {Pati{\~n}o-{\'A}lvarez} V.~M.,  {Amaya-Almaz{\'a}n} R.~A.,
  {Carrasco} L.,  2020, \mn@doi [\apj] {10.3847/1538-4357/ab6ef6}, \href
  {https://ui.adsabs.harvard.edu/abs/2020ApJ...891...68C} {891, 68}

\bibitem[\protect\citeauthoryear{{Chiang} \& {B{\"o}ttcher}}{{Chiang} \&
  {B{\"o}ttcher}}{2002}]{Chiang_2002}
{Chiang} J.,  {B{\"o}ttcher} M.,  2002, \mn@doi [\apj] {10.1086/324294}, \href
  {https://ui.adsabs.harvard.edu/abs/2002ApJ...564...92C} {564, 92}

\bibitem[\protect\citeauthoryear{{Coppi} \& {Aharonian}}{{Coppi} \&
  {Aharonian}}{1999}]{Coppi_1999}
{Coppi} P.~S.,  {Aharonian} F.~A.,  1999, \mn@doi [\apjl] {10.1086/312168},
  \href {https://ui.adsabs.harvard.edu/abs/1999ApJ...521L..33C} {521, L33}

\bibitem[\protect\citeauthoryear{{Costamante}, {Cutini}, {Tosti}, {Antolini}
  \& {Tramacere}}{{Costamante} et~al.}{2018}]{costamate18}
{Costamante} L.,  {Cutini} S.,  {Tosti} G.,  {Antolini} E.,   {Tramacere} A.,
  2018, \mn@doi [\mnras] {10.1093/mnras/sty887}, \href
  {https://ui.adsabs.harvard.edu/abs/2018MNRAS.477.4749C} {477, 4749}

\bibitem[\protect\citeauthoryear{{D'arcangelo} et~al.,}{{D'arcangelo}
  et~al.}{2009}]{D_arcangelo_2009}
{D'arcangelo} F.~D.,  et~al., 2009, \mn@doi [\apj]
  {10.1088/0004-637X/697/2/985}, \href
  {https://ui.adsabs.harvard.edu/abs/2009ApJ...697..985D} {697, 985}

\bibitem[\protect\citeauthoryear{{Dermer}, {Sturner}  \&
  {Schlickeiser}}{{Dermer} et~al.}{1997}]{Dermer_1997}
{Dermer} C.~D.,  {Sturner} S.~J.,   {Schlickeiser} R.,  1997, \mn@doi [\apjs]
  {10.1086/312972}, \href
  {https://ui.adsabs.harvard.edu/abs/1997ApJS..109..103D} {109, 103}

\bibitem[\protect\citeauthoryear{{Donea} \& {Protheroe}}{{Donea} \&
  {Protheroe}}{2003}]{donea2003}
{Donea} A.-C.,  {Protheroe} R.~J.,  2003, \mn@doi [Astroparticle Physics]
  {10.1016/S0927-6505(02)00155-X}, \href
  {https://ui.adsabs.harvard.edu/abs/2003APh....18..377D} {18, 377}

\bibitem[\protect\citeauthoryear{{Dotson}, {Georganopoulos}, {Kazanas}  \&
  {Perlman}}{{Dotson} et~al.}{2012}]{Dotson_2012}
{Dotson} A.,  {Georganopoulos} M.,  {Kazanas} D.,   {Perlman} E.~S.,  2012,
  \mn@doi [\apjl] {10.1088/2041-8205/758/1/L15}, \href
  {https://ui.adsabs.harvard.edu/abs/2012ApJ...758L..15D} {758, L15}

\bibitem[\protect\citeauthoryear{{Dotson}, {Georganopoulos}, {Meyer}  \&
  {McCann}}{{Dotson} et~al.}{2015}]{Dotson_2015}
{Dotson} A.,  {Georganopoulos} M.,  {Meyer} E.~T.,   {McCann} K.,  2015,
  \mn@doi [\apj] {10.1088/0004-637X/809/2/164}, \href
  {https://ui.adsabs.harvard.edu/abs/2015ApJ...809..164D} {809, 164}

\bibitem[\protect\citeauthoryear{{Dutka} et~al.,}{{Dutka}
  et~al.}{2017}]{dutka17}
{Dutka} M.~S.,  et~al., 2017, \mn@doi [\apj] {10.3847/1538-4357/835/2/182},
  \href {https://ui.adsabs.harvard.edu/abs/2017ApJ...835..182D} {835, 182}

\bibitem[\protect\citeauthoryear{{Ghisellini} \& {Madau}}{{Ghisellini} \&
  {Madau}}{1996}]{ghisselini_96}
{Ghisellini} G.,  {Madau} P.,  1996, \mn@doi [\mnras] {10.1093/mnras/280.1.67},
  \href {https://ui.adsabs.harvard.edu/abs/1996MNRAS.280...67G} {280, 67}

\bibitem[\protect\citeauthoryear{{Ghisellini}, {Tavecchio}  \&
  {Chiaberge}}{{Ghisellini} et~al.}{2005}]{ghisselini2005}
{Ghisellini} G.,  {Tavecchio} F.,   {Chiaberge} M.,  2005, \mn@doi [\aap]
  {10.1051/0004-6361:20041404}, \href
  {https://ui.adsabs.harvard.edu/abs/2005A&A...432..401G} {432, 401}

\bibitem[\protect\citeauthoryear{{Ghisellini}, {Tavecchio}, {Foschini},
  {Ghirlanda}, {Maraschi}  \& {Celotti}}{{Ghisellini}
  et~al.}{2010}]{Ghisselini_2010}
{Ghisellini} G.,  {Tavecchio} F.,  {Foschini} L.,  {Ghirlanda} G.,  {Maraschi}
  L.,   {Celotti} A.,  2010, \mn@doi [\mnras]
  {10.1111/j.1365-2966.2009.15898.x}, \href
  {https://ui.adsabs.harvard.edu/abs/2010MNRAS.402..497G} {402, 497}

\bibitem[\protect\citeauthoryear{Ghisellini, Tavecchio, Foschini  \&
  Ghirlanda}{Ghisellini et~al.}{2011}]{ghisselini2011}
Ghisellini G.,  Tavecchio F.,  Foschini L.,   Ghirlanda G.,  2011, \mn@doi
  [Monthly Notices of the Royal Astronomical Society]
  {10.1111/j.1365-2966.2011.18578.x}, 414, 2674

\bibitem[\protect\citeauthoryear{{Hartman} et~al.,}{{Hartman}
  et~al.}{2001}]{hartman2001}
{Hartman} R.~C.,  et~al., 2001, \mn@doi [\apj] {10.1086/320970}, \href
  {https://ui.adsabs.harvard.edu/abs/2001ApJ...553..683H} {553, 683}

\bibitem[\protect\citeauthoryear{{Harvey}, {Georganopoulos}  \&
  {Meyer}}{{Harvey} et~al.}{2020}]{georganopoulos2020}
{Harvey} A. L.~W.,  {Georganopoulos} M.,   {Meyer} E.~T.,  2020, \mn@doi
  [Nature Communications] {10.1038/s41467-020-19296-6}, \href
  {https://ui.adsabs.harvard.edu/abs/2020NatCo..11.5475H} {11, 5475}

\bibitem[\protect\citeauthoryear{{Hayashida} et~al.,}{{Hayashida}
  et~al.}{2012}]{hayashida12}
{Hayashida} M.,  et~al., 2012, \mn@doi [\apj] {10.1088/0004-637X/754/2/114},
  \href {https://ui.adsabs.harvard.edu/abs/2012ApJ...754..114H} {754, 114}

\bibitem[\protect\citeauthoryear{{Hovatta} \& {Lindfors}}{{Hovatta} \&
  {Lindfors}}{2019}]{hovatta19}
{Hovatta} T.,  {Lindfors} E.,  2019, \mn@doi [\nar]
  {10.1016/j.newar.2020.101541}, \href
  {https://ui.adsabs.harvard.edu/abs/2019NewAR..8701541H} {87, 101541}

\bibitem[\protect\citeauthoryear{{Impey} \& {Neugebauer}}{{Impey} \&
  {Neugebauer}}{1988}]{impey_1988}
{Impey} C.~D.,  {Neugebauer} G.,  1988, \mn@doi [\aj] {10.1086/114638}, \href
  {https://ui.adsabs.harvard.edu/abs/1988AJ.....95..307I} {95, 307}

\bibitem[\protect\citeauthoryear{{Isler} et~al.,}{{Isler}
  et~al.}{2013}]{Isler_2013}
{Isler} J.~C.,  et~al., 2013, \mn@doi [\apj] {10.1088/0004-637X/779/2/100},
  \href {https://ui.adsabs.harvard.edu/abs/2013ApJ...779..100I} {779, 100}

\bibitem[\protect\citeauthoryear{{Isler} et~al.,}{{Isler}
  et~al.}{2015}]{isler2015}
{Isler} J.~C.,  et~al., 2015, \mn@doi [\apj] {10.1088/0004-637X/804/1/7}, \href
  {https://ui.adsabs.harvard.edu/abs/2015ApJ...804....7I} {804, 7}

\bibitem[\protect\citeauthoryear{{Jorstad} et~al.,}{{Jorstad}
  et~al.}{2013}]{jorstad13}
{Jorstad} S.~G.,  et~al., 2013, \mn@doi [\apj] {10.1088/0004-637X/773/2/147},
  \href {https://ui.adsabs.harvard.edu/abs/2013ApJ...773..147J} {773, 147}

\bibitem[\protect\citeauthoryear{{Kang}, {Chen}  \& {Wu}}{{Kang}
  et~al.}{2014}]{kang2014}
{Kang} S.-J.,  {Chen} L.,   {Wu} Q.,  2014, \mn@doi [\apjs]
  {10.1088/0067-0049/215/1/5}, \href
  {https://ui.adsabs.harvard.edu/abs/2014ApJS..215....5K} {215, 5}

\bibitem[\protect\citeauthoryear{{Kramarenko}, {Pushkarev}, {Kovalev},
  {Lister}, {Hovatta}  \& {Savolainen}}{{Kramarenko}
  et~al.}{2022}]{Kramarenko2022}
{Kramarenko} I.~G.,  {Pushkarev} A.~B.,  {Kovalev} Y.~Y.,  {Lister} M.~L.,
  {Hovatta} T.,   {Savolainen} T.,  2022, \mn@doi [\mnras]
  {10.1093/mnras/stab3358}, \href
  {https://ui.adsabs.harvard.edu/abs/2022MNRAS.510..469K} {510, 469}

\bibitem[\protect\citeauthoryear{{Lazzati}, {Ghisellini}  \&
  {Celotti}}{{Lazzati} et~al.}{1999}]{lazzati99}
{Lazzati} D.,  {Ghisellini} G.,   {Celotti} A.,  1999, \mn@doi [\mnras]
  {10.1046/j.1365-8711.1999.02970.x}, \href
  {https://ui.adsabs.harvard.edu/abs/1999MNRAS.309L..13L} {309, L13}

\bibitem[\protect\citeauthoryear{{Le{\'o}n-Tavares} et~al.,}{{Le{\'o}n-Tavares}
  et~al.}{2013}]{tavares13}
{Le{\'o}n-Tavares} J.,  et~al., 2013, \mn@doi [\apjl]
  {10.1088/2041-8205/763/2/L36}, \href
  {https://ui.adsabs.harvard.edu/abs/2013ApJ...763L..36L} {763, L36}

\bibitem[\protect\citeauthoryear{{Liodakis}, {Romani}, {Filippenko}, {Kocevski}
   \& {Zheng}}{{Liodakis} et~al.}{2019}]{Liodakis_2019}
{Liodakis} I.,  {Romani} R.~W.,  {Filippenko} A.~V.,  {Kocevski} D.,   {Zheng}
  W.,  2019, \mn@doi [\apj] {10.3847/1538-4357/ab26b7}, \href
  {https://ui.adsabs.harvard.edu/abs/2019ApJ...880...32L} {880, 32}

\bibitem[\protect\citeauthoryear{{MacDonald}, {Marscher}, {Jorstad}  \&
  {Joshi}}{{MacDonald} et~al.}{2015}]{macdonald15}
{MacDonald} N.~R.,  {Marscher} A.~P.,  {Jorstad} S.~G.,   {Joshi} M.,  2015,
  \mn@doi [\apj] {10.1088/0004-637X/804/2/111}, \href
  {https://ui.adsabs.harvard.edu/abs/2015ApJ...804..111M} {804, 111}

\bibitem[\protect\citeauthoryear{{MacDonald}, {Jorstad}  \&
  {Marscher}}{{MacDonald} et~al.}{2017}]{macdonald17}
{MacDonald} N.~R.,  {Jorstad} S.~G.,   {Marscher} A.~P.,  2017, \mn@doi [\apj]
  {10.3847/1538-4357/aa92c8}, \href
  {https://ui.adsabs.harvard.edu/abs/2017ApJ...850...87M} {850, 87}

\bibitem[\protect\citeauthoryear{{Majumder}, {Mitra}, {Chatterjee}, {Urry},
  {Bailyn}  \& {Nandi}}{{Majumder} et~al.}{2019}]{majumdar_19}
{Majumder} A.,  {Mitra} K.,  {Chatterjee} R.,  {Urry} C.~M.,  {Bailyn} C.~D.,
  {Nandi} P.,  2019, \mn@doi [\mnras] {10.1093/mnras/stz2557}, \href
  {https://ui.adsabs.harvard.edu/abs/2019MNRAS.490..124M} {490, 124}

\bibitem[\protect\citeauthoryear{{Maraschi}, {Ghisellini}  \&
  {Celotti}}{{Maraschi} et~al.}{1992}]{maraschi1992}
{Maraschi} L.,  {Ghisellini} G.,   {Celotti} A.,  1992, \mn@doi [\apjl]
  {10.1086/186531}, \href
  {https://ui.adsabs.harvard.edu/abs/1992ApJ...397L...5M} {397, L5}

\bibitem[\protect\citeauthoryear{{Marscher}}{{Marscher}}{2005}]{marscher_2005}
{Marscher} A.~P.,  2005, in {Romney} J.,  {Reid} M.,  eds,  Astronomical
  Society of the Pacific Conference Series Vol. 340, Future Directions in High
  Resolution Astronomy. p.~25

\bibitem[\protect\citeauthoryear{{Marscher}}{{Marscher}}{2016}]{marscher2016}
{Marscher} A.,  2016, \mn@doi [Galaxies] {10.3390/galaxies4040037}, \href
  {https://ui.adsabs.harvard.edu/abs/2016Galax...4...37M} {4, 37}

\bibitem[\protect\citeauthoryear{{Marscher} \& {Gear}}{{Marscher} \&
  {Gear}}{1985}]{marscher1985}
{Marscher} A.~P.,  {Gear} W.~K.,  1985, \mn@doi [\apj] {10.1086/163592}, \href
  {https://ui.adsabs.harvard.edu/abs/1985ApJ...298..114M} {298, 114}

\bibitem[\protect\citeauthoryear{{Marscher} et~al.,}{{Marscher}
  et~al.}{2008}]{marscher2008}
{Marscher} A.~P.,  et~al., 2008, \mn@doi [\nat] {10.1038/nature06895}, \href
  {https://ui.adsabs.harvard.edu/abs/2008Natur.452..966M} {452, 966}

\bibitem[\protect\citeauthoryear{{Marscher} et~al.,}{{Marscher}
  et~al.}{2010}]{marscher2010}
{Marscher} A.~P.,  et~al., 2010, \mn@doi [\apjl]
  {10.1088/2041-8205/710/2/L126}, \href
  {https://ui.adsabs.harvard.edu/abs/2010ApJ...710L.126M} {710, L126}

\bibitem[\protect\citeauthoryear{{Meyer}, {Fossati}, {Georganopoulos}  \&
  {Lister}}{{Meyer} et~al.}{2012}]{Meyer_2012}
{Meyer} E.~T.,  {Fossati} G.,  {Georganopoulos} M.,   {Lister} M.~L.,  2012,
  \mn@doi [\apjl] {10.1088/2041-8205/752/1/L4}, \href
  {https://ui.adsabs.harvard.edu/abs/2012ApJ...752L...4M} {752, L4}

\bibitem[\protect\citeauthoryear{{Nalewajko}, {Begelman}  \&
  {Sikora}}{{Nalewajko} et~al.}{2014}]{nalewajko14}
{Nalewajko} K.,  {Begelman} M.~C.,   {Sikora} M.,  2014, \mn@doi [\apj]
  {10.1088/0004-637X/789/2/161}, \href
  {https://ui.adsabs.harvard.edu/abs/2014ApJ...789..161N} {789, 161}

\bibitem[\protect\citeauthoryear{{Prince}, {Gupta}  \& {Nalewajko}}{{Prince}
  et~al.}{2019}]{prince2019}
{Prince} R.,  {Gupta} N.,   {Nalewajko} K.,  2019, \mn@doi [\apj]
  {10.3847/1538-4357/ab3afa}, \href
  {https://ui.adsabs.harvard.edu/abs/2019ApJ...883..137P} {883, 137}

\bibitem[\protect\citeauthoryear{{Pushkarev}, {Hovatta}, {Kovalev}, {Lister},
  {Lobanov}, {Savolainen}  \& {Zensus}}{{Pushkarev}
  et~al.}{2012}]{pushkarev2012}
{Pushkarev} A.~B.,  {Hovatta} T.,  {Kovalev} Y.~Y.,  {Lister} M.~L.,  {Lobanov}
  A.~P.,  {Savolainen} T.,   {Zensus} J.~A.,  2012, \mn@doi [\aap]
  {10.1051/0004-6361/201219173}, \href
  {https://ui.adsabs.harvard.edu/abs/2012A&A...545A.113P} {545, A113}

\bibitem[\protect\citeauthoryear{{Rani} et~al.,}{{Rani}
  et~al.}{2018}]{Rani2018}
{Rani} B.,  et~al., 2018, \mn@doi [\apj] {10.3847/1538-4357/aab785}, \href
  {https://ui.adsabs.harvard.edu/abs/2018ApJ...858...80R} {858, 80}

\bibitem[\protect\citeauthoryear{{Roy}, {Chatterjee}, {Joshi}  \&
  {Ghosh}}{{Roy} et~al.}{2019}]{namrata_19}
{Roy} N.,  {Chatterjee} R.,  {Joshi} M.,   {Ghosh} A.,  2019, \mn@doi [\mnras]
  {10.1093/mnras/sty2748}, \href
  {https://ui.adsabs.harvard.edu/abs/2019MNRAS.482..743R} {482, 743}

\bibitem[\protect\citeauthoryear{{Roy}, {Patel}, {Sarkar}, {Chatterjee}  \&
  {Chitnis}}{{Roy} et~al.}{2021}]{roy2021}
{Roy} A.,  {Patel} S.~R.,  {Sarkar} A.,  {Chatterjee} A.,   {Chitnis} V.~R.,
  2021, \mn@doi [\mnras] {10.1093/mnras/stab975}, \href
  {https://ui.adsabs.harvard.edu/abs/2021MNRAS.504.1103R} {504, 1103}

\bibitem[\protect\citeauthoryear{{Saito}, {Stawarz}, {Tanaka}, {Takahashi},
  {Madejski}  \& {D'Ammando}}{{Saito} et~al.}{2013}]{Saito_2013}
{Saito} S.,  {Stawarz} {\L}.,  {Tanaka} Y.~T.,  {Takahashi} T.,  {Madejski} G.,
    {D'Ammando} F.,  2013, \mn@doi [\apjl] {10.1088/2041-8205/766/1/L11}, \href
  {https://ui.adsabs.harvard.edu/abs/2013ApJ...766L..11S} {766, L11}

\bibitem[\protect\citeauthoryear{{Sikora}, {Begelman}  \& {Rees}}{{Sikora}
  et~al.}{1994}]{sikora_1994}
{Sikora} M.,  {Begelman} M.~C.,   {Rees} M.~J.,  1994, \mn@doi [\apj]
  {10.1086/173633}, \href
  {https://ui.adsabs.harvard.edu/abs/1994ApJ...421..153S} {421, 153}

\bibitem[\protect\citeauthoryear{{Sikora}, {Stawarz}, {Moderski}, {Nalewajko}
  \& {Madejski}}{{Sikora} et~al.}{2009}]{Sikora_2009}
{Sikora} M.,  {Stawarz} {\L}.,  {Moderski} R.,  {Nalewajko} K.,   {Madejski}
  G.~M.,  2009, \mn@doi [\apj] {10.1088/0004-637X/704/1/38}, \href
  {https://ui.adsabs.harvard.edu/abs/2009ApJ...704...38S} {704, 38}

\bibitem[\protect\citeauthoryear{{Tavecchio} \& {Mazin}}{{Tavecchio} \&
  {Mazin}}{2009}]{tavecchio2009}
{Tavecchio} F.,  {Mazin} D.,  2009, \mn@doi [\mnras]
  {10.1111/j.1745-3933.2008.00584.x}, \href
  {https://ui.adsabs.harvard.edu/abs/2009MNRAS.392L..40T} {392, L40}

\bibitem[\protect\citeauthoryear{{Urry} \& {Mushotzky}}{{Urry} \&
  {Mushotzky}}{1982}]{urry_1982}
{Urry} C.~M.,  {Mushotzky} R.~F.,  1982, \mn@doi [\apj] {10.1086/159607}, \href
  {https://ui.adsabs.harvard.edu/abs/1982ApJ...253...38U} {253, 38}

\bibitem[\protect\citeauthoryear{{Urry} \& {Padovani}}{{Urry} \&
  {Padovani}}{1995}]{urry1995}
{Urry} C.~M.,  {Padovani} P.,  1995, \mn@doi [\pasp] {10.1086/133630}, \href
  {https://ui.adsabs.harvard.edu/abs/1995PASP..107..803U} {107, 803}

\bibitem[\protect\citeauthoryear{{Valtaoja}, {L{\"a}hteenm{\"a}ki},
  {Ter{\"a}sranta}  \& {Lainela}}{{Valtaoja} et~al.}{1999}]{valtoja99}
{Valtaoja} E.,  {L{\"a}hteenm{\"a}ki} A.,  {Ter{\"a}sranta} H.,   {Lainela} M.,
   1999, \mn@doi [\apjs] {10.1086/313170}, \href
  {https://ui.adsabs.harvard.edu/abs/1999ApJS..120...95V} {120, 95}

\bibitem[\protect\citeauthoryear{{Whiting}}{{Whiting}}{2005}]{whitting2005}
{Whiting} M.~T.,  2005, \memsai, \href
  {https://ui.adsabs.harvard.edu/abs/2005MmSAI..76...61W} {76, 61}

\makeatother
\end{thebibliography}
\end{document}